# Dual photoisomerization mechanism of azobenzene embedded in a lipid membrane

Silvio Osella[1,2]*, Giovanni Granucci[3], Maurizio Persico[3], Stefan Knippenberg[4,5]*

[1] Chemical and Biological Systems Simulation Lab, Centre of New Technologies, University of Warsaw, Banacha 2C, 02-097 Warsaw, Poland.

[2] Materials and Process Simulation Center (MSC), California Institute of Technology, MC 139-74, Pasadena CA, 91125, USA.

[3] Dipartimento di Chimica e Chimica Industriale, Universitá di Pisa, v. Moruzzi 13, I-56124 Pisa, Italy.

[4] Hasselt University, Theory Lab, Agoralaan Building D, 3590 Diepenbeek, Belgium.

[5] Université Libre de Bruxelles, Spectroscopy, Quantum Chemistry and Atmospheric Remote Sensing (SQUARES), 50 Avenue F. Roosevelt, C.P. 160/09, B-1050 Brussels, Belgium.

Email addresses: silvio.osella@cent.uw.edu.pl (SO); stefan.knippenberg@uhasselt.be (SK)

**Abstract**

The photoisomerization of chromophores embedded in biological environments is of high importance for biomedical applications, but it is still challenging to define the photoisomerization mechanism both experimentally and computationally. We present here a computational study of the azobenzene molecule embedded in a DPPC lipid membrane, and assess the photoisomerization mechanism by means of the quantum mechanics/molecular mechanics surface hopping (QM/MM-SH) method. We observe that while the trans-to-cis isomerization is a slow process governed by a torsional mechanism due to the strong interaction with the environment, the cis-to-trans mechanism is completed in sub-ps time scale and is governed by a pedal-like mechanism in which both weaker interactions with the environment and a different geometry of the potential energy surface play a key role.

# 1. Introduction

Photochromic molecules possess the ability to undergo reversible isomerization triggered by light, between stable and metastable states. As a result, many different classes of molecules which undergo isomerization have been designed and synthetized over the years. At least five different classes of isomerization processes can be considered:[1] (1) trans/cis (or E/Z) isomerization such as stilbene and azobenzene, (2) pericyclic reactions like spiropyranes/oxazines, (3) [1,3] sigmatropic shifts, (4) photoelectrocyclic ring openings like for diarylethenes and (5) intramolecular hydrogen transfer such as for quinones.[2,3]

One of the most used molecules falling in the first class is azobenzene (AZO), which – upon absorption of UV light – reversibly switches between the planar trans state and the non-planar metastable cis isomer, inducing a trans-to-cis conversion. The opposite cis-to-trans process can either be triggered by visible light or by thermal energy. Despite the wide use of the azobenzene molecule in organic field effect transistors (OFET)[4] and in DNA/RNA related applications,[5] its use as photochromic probe in lipid membranes is still not fully exploited. Commonly, the azobenzene molecule is considered as membrane marker, in which it is attached to one of the membrane lipid tails forming so called photoswitchable phospholipids,[6] and its photoisomerization is considered for the ability of modifying the shape, fluidity and permeability of the membrane thanks to the increase in thickness arising from the trans-to-cis isomerization.[7,8] Recently, Lanzani *et al.* reported a study on photophysical properties of an azobenzene derivative embedded in a E.coli membrane, showing that the photoisomerization ability of the probe strongly depends on the viscosity of the environment.[9]

Yet, the study of the photochromic properties of azobenzene as probe for membrane phase detection is still not considered. This is an important area of research, since the phase of a membrane bilayer determines if the cell is healthy or not. In particular, a change in lipid membrane phase from a stiff one to a depleted, liquid disordered one is a key parameter to discriminate between cancerogenic and healthy cells.[10–13] This is an extremely important issue, and the study of the optical response of a probe such as azobenzene in a biological environment can give strong insight into the phase of the membrane, especially due to the different response of the two isomers. Thanks to the ability of azobenzene to undergo photoswitching,[14] we can consider one state as active and the other as inactive to fluorescence, giving a dual 'on/off' response in which the only a priori knowledge needed is the starting isomerization state of the probe.[15,16] In this way, if one isomer is active in a specific phase and the other in a different one, it will be possible to screen for different membrane compositions.

In the current study, we consider the photochromic properties of azobenzene when embedded in a liquid disordered (Ld) membrane phase, to assess the hypothesis that azobenzene can act as an 'on/off' molecular switch whose fluorescent properties depend on the state of the probe (trans or cis). A lipid bilayer of DiPalmitoylPhosphatidylCholine (DPPC) at 323K is here used to avoid complications due to eventual interactions of the probe with spurious unsaturation of the tails. Due to the large difference in timescale for the probe photoisomerization (fs to ps) and the environment response (hundreds of ns to μs), complementary computational methods are required to simulate the whole process. As a result, we will resort to a multiscale computational approach to elucidate the photoswitching mechanism for both trans-to-cis and cis-to-trans processes. To such an aim, we use molecular dynamic (MD) simulations to equilibrate the position of the probe in the membrane, and subsequently focus on the photoswitch mechanism using hybrid quantum mechanics / molecular mechanics (QM/MM) simulations coupled with Surface Hopping (QM/MM-SH) dynamics, in order to obtain a detailed understanding of the complex isomerization process taking place in the lipid membrane.

Our study confirms the presence of two photoisomerization mechanisms depending on the starting isomer considered, which in turn affects the optical response of the probe, supporting the dual 'on/off' ability of the azobenzene probe.

## 2. Methodology

In order to obtain the input structures needed for the QM/MM-SH simulations, molecular dynamics (MD) simulations were performed for both trans and cis isomers in a DPPC membrane at 323K. The OPLS-AA force field was used to describe both the DPPC membrane and the AZO isomers (structures are reported in SI). For the latter, force field parameters as reported in literature[17] were considered, and the partial atomic charges were obtained at the B3LYP/cc-pVTZ level of theory, using the Gaussian 16 software (see SI for more details).[18] Periodic boundary conditions were used along the whole study, with an integration time step of 2 fs using LINCS constraints, Particle Mesh Ewald (PME) Coulomb[19] and van der Waals interactions with 1.2 nm cutoff and dispersion correction in an NPT ensemble (Nose-Hoover thermostat with time constant of 0.4 ps and semi-isotropic Parrinello-Rahman barostat with time constant of 1 ps and a compressibility of $4.5 \times 10^{-5}$ bar$^{-1}$).[20–22] The system was solvated with water, using TIP3P parameters. The membrane surface has been oriented to be parallel to the xy plane.

The complete simulation protocol used in this study is the following:

1. Equilibration of the solvated model DPPC membrane consisting out of 64 molecule per layer. An annealing procedure has been considered, starting with a short MD of 10 ns at 323K, followed by 50 ns at 500 K and final cooling at 323K for 100 ns.
2. Ground state MD of the full system with AZO. One AZO molecule (either trans or cis) has been inserted in the water phase of the equilibrated DPPC membrane, and a 500 ns long MD at 323K has been performed.
3. The last frame of the previous MD has been considered as starting point for the QM/MM simulation. A cylindrical portion of the infinite membrane (AZO isomer, 47 lipids and 122 water molecules within 1 nm from the AZO) is selected for the QM/MM analysis, in which the AZO molecule is described by the floating occupation molecular orbitals-configuration interaction (FOMO-CI) semiempirical method[23–26] and the rest is treated at the molecular mechanics (MM) level with electrostatic embedding (see Figure 1 for the QM/MM partition). An additional equilibration is run for 10 ps, with timestep of 0.1 fs. To avoid the disassembly of the membrane, the methyl groups and the P atoms of the membrane molecules at the border have been constrained.
4. From the equilibrated system, the AZO molecule is vertically excited to the $S_1$ (n → π*) state at phase space points of the QM/MM trajectory selected on the basis of dipole transition probabilities for a total of 298 initial trajectories for the trans isomer (AZO-t) and 397 for the cis isomer (AZO-c). The $S_1$ state has been chosen as the starting state for the ab initio MD accordingly to reported literature[17,27] which show an extremely fast decay (in the fs time window) from higher states ($S_2$-$S_5$) to the selected one, making it the ideal candidate for the photoisomerization study.

5. QM/MM-SH nonadiabatic trajectories with quantum decoherence corrections are then run for 10 ps with timestep of 0.1 fs, starting from the initial condition of (4). A detailed description of the method can be found in Refs 23-26 and in the SI.

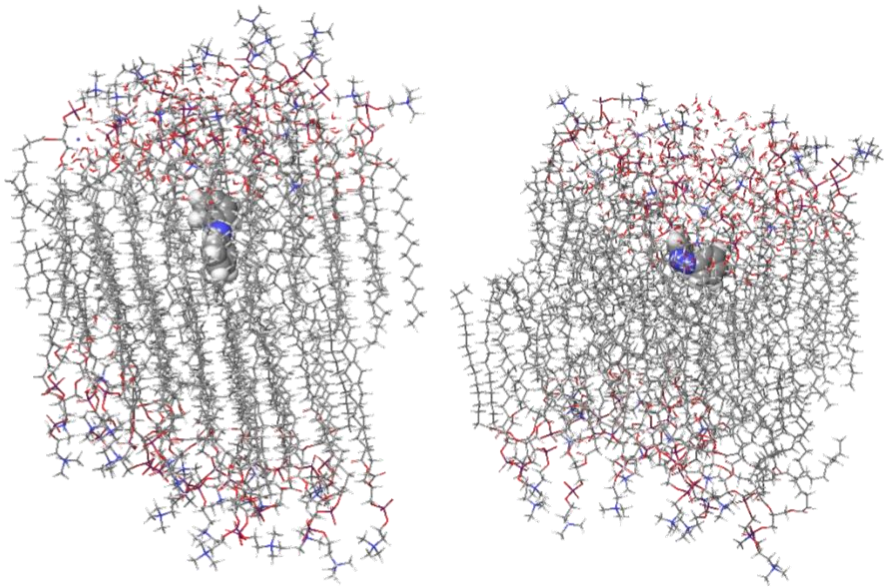

**Figure 1. Input structures for AZO-t (left) and AZO-c (right) isomers in DPPC membrane, considered for the QM/MM simulations. Color code: gray-carbon; blue-nitrogen; red-oxygen; white-hydrogen; orange-phosphorous.**

## 3. Results and Discussion

### 3.1 MD simulations

The transition dipole moment (TDM) analysis for both isomers reveals a different orientation for AZO-t and AZO-c, with the former lying in an almost parallel orientation at an angle of around 20 degrees with respect to the membrane tails, and the latter with a perpendicular orientation and an angle around 82 degrees (Figure 2a). A minor population of trans is found with its TDM parallel to the membrane surface. This particular orientation can be compared to the one of the diphenylhexatriene probe, of which the minor population parallel to the surface of the liquid (dis)ordered lipid bilayers was seen both in experiment and in multiscale modeling.[28–30] The angle between the TDM and the membrane normal is used to check the equilibration of the simulated lipid bilayer. For the specific case, the trans isomer can be considered equilibrated after 300 ns, while the cis after 200 ns (all the reported analyses consider the 300-500 ns time window, see Figure S2 in Supplementary Information). The TDM analysis shows that AZO-t is oriented almost parallel to the membrane normal, with an angle of around 20-30 degrees, while for AZO-c the angle is peaking around 80 degrees, leading to an orientation in which the phenyl rings are parallel to the plane of the DPPC tails (see Figure 2a and S1). Despite the difference in orientation, the position of the AZO molecule in the membrane is similar, with the trans isomer positioned slightly deeper in the membrane, at 0.95 nm from the bilayer center, and the cis isomer closer to the membrane head group region, at a distance of 1.15 nm (Figure 2b).

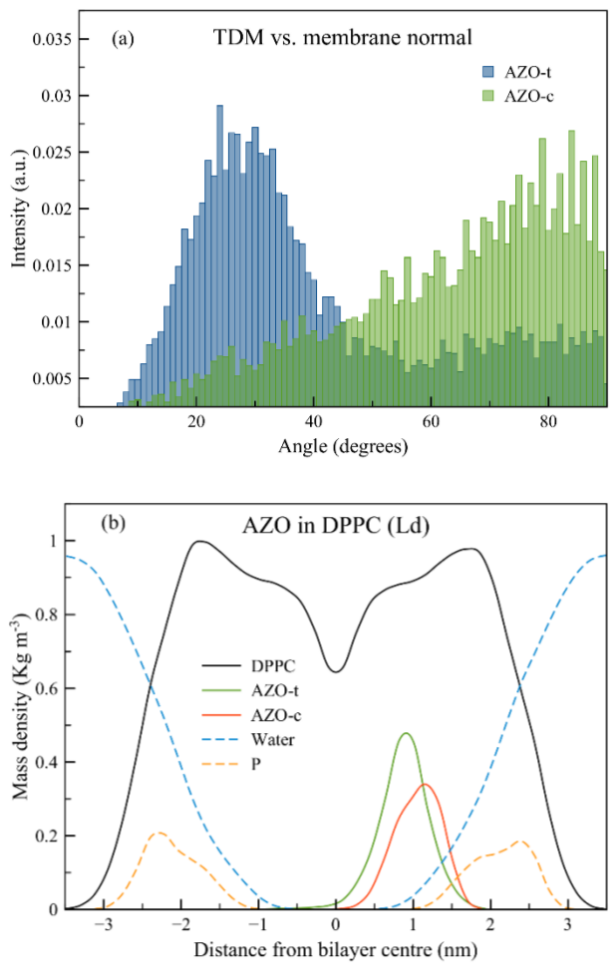

**Figure 2. (a) Angle between the TDM and the membrane normal. TDM is considered as the vector between the two C atoms in para to the N=N bond for AZO-t, and as the C-N vector making an angle of 63 degrees with the N=N bond for AZO-c. (b) Density plot of the position of the two azobenzene isomers in the DPPC membrane. P refers to the phosphorous atoms of the membrane head.**

A final geometrical analysis of the CNNC dihedral angle (the dihedral angle formed by the N=N bond and the adjacent C atoms) proves that there is no crossing of conformations from AZO-t to AZO-c and *vice versa* during the simulation time, as a narrow distribution of the CNNC dihedral angle of 174.3 and 6 degrees has been found for AZO-t and AZO-c, respectively (Figure S3 in SI). In addition, the analysis of both the CNN planar angles reveals that while both CNN1 and CNN2 have the same values, they differ for the two isomers, with angles of 112 degrees for AZO-t and of 125 degrees for AZO-c, respectively.

Finally, to assess the quality of the force field and the validity of the protocol used, we computed the photoselection ability of the two probes in the membrane (we assume that the incident beam has a polarization parallel to the membrane normal). Analyzing the $\cos^2$ of the angle between the TDM and the z-axis, the photoselection is almost twice as probable for the AZO-t isomer (0.56) than the AZO-c (0.25), in agreement with our previous study.[16]

### 3.2. QM/MM-SH simulations

In this section, we first analyze the electronic state population and estimate the isomerization quantum yields from the QM/MM-SH simulations. Then, we describe the isomerization mechanism and the fluorescence properties, in which the role of the environment is discussed.

#### 3.2.1 Photodynamic population and quantum yield

From the ground state equilibration, we got an average dihedral value of 178.5 degrees for AZO-t and 7 degrees for AZO-c (considering the [0,180] range). From the 10 ps QM/MM equilibration of both isomers in the DPPC membrane, we extracted the absorption spectra, which are reported in Figure 3.

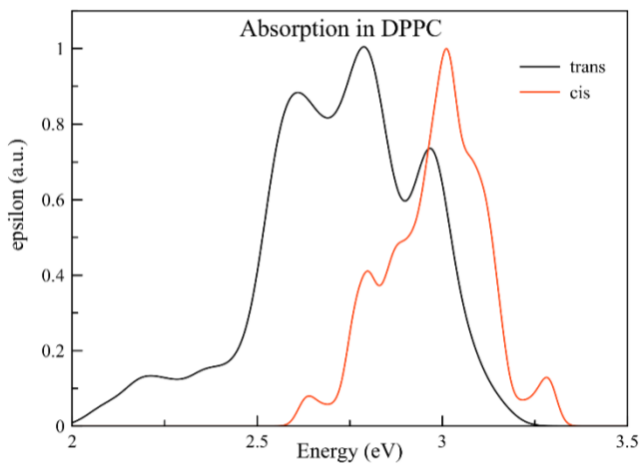

**Figure 3. Absorption spectra of the trans and cis azobenzene isomers in DPPC.**

For this analysis the first five low-lying excited states have been considered. The first interesting result to note is that the band of the AZO-t for the $S_1$ n-$\pi$* transition is much more intense (oscillator strength of 0.13) compared to the case in vacuum[31] and has been computed at an energy of 2.2 eV. The same transition for the AZO-c has been found at 2.6 eV (oscillator strength of 0.08), in agreement with data of azobenzene embedded in different solvents.[32–34] Moreover, the red shift of the AZO-t of 0.5 eV, is in full agreement with our previous calculations of the same probe embedded in a DOPC membrane.[16] As expected, the most intense absorption peak is related to the $S_2$ $\pi$-$\pi$* transition, located at 2.6 eV for AZO-t and 3 eV for AZO-c.[35]

The key results obtained from the surface hopping simulations are reported in Figure 4. The depicted population decay $P$ can be fitted to an approximate first-order kinetic equation with

$$P_1(t) = 1 \text{ for } t < t_d,$$

$$P_1(t) = \exp\left(-(t - t_d)/\tau\right) \quad \text{for } t \geq t_d \quad \text{and}$$

$$P_0(t) = 1 - P_1(t),$$

where $t_d$ denotes a delay time. The results are reported in Table 1, for both the gas phase and DPPC environment, for the active trajectories (for the inactive ones, see Table S1).

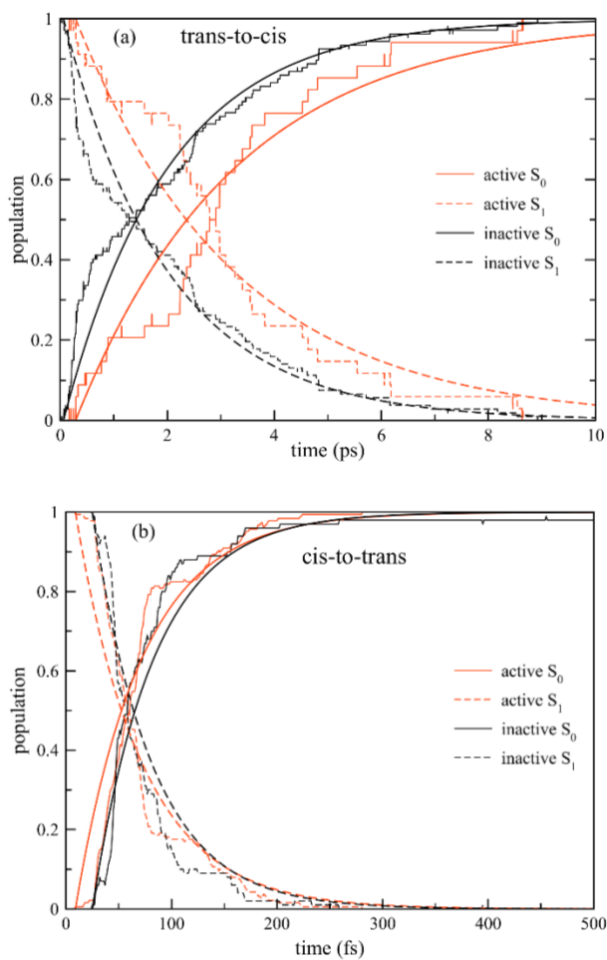

**Figure 4.** Population analysis of the ground state ($S_0$) and first excited state ($S_1$). For the trans-to-cis process (a), active trajectories finished in cis, inactive stayed in the trans conformation while for the cis-to-trans process (b) active trajectories finished in trans, inactive stayed in the cis conformation.

**Table 1.** Decay times from exponential fit of population decay time for $S_0$ ($\tau_0$) and for the $S_1$ state ($\tau_1$), and photoisomerization quantum yield $\Phi$ for the active trajectories. Times are reported in ps.

| environment | isomerization | $\tau_0$ | $\tau_1$ | $\Phi$ |
|---|---|---|---|---|
| **vacuum** | Trans-cis | 0.32 | 0.31 | 0.35 |
| | Cis-trans | 0.06 | 0.06 | 0.50 |
| **DPPC 323K** | Trans-cis | 3.00 | 2.98 | 0.24 |
| | Cis-trans | 0.06 | 0.06 | 0.65 |

Making use of exponential fits of these graphs to obtain characteristic times, the trans-to-cis photoisomerization in DPPC environment is tenfold slowed down compared to the gas phase, from 0.3 ps in gas to 3.0 ps in the environment, as well as reported lifetimes values in different solvents such as methanol and ethylene glycol.[5,17] By contrast, the reverse process from cis-to-trans is almost unaffected by the presence of the lipid membrane, and is concluded in 0.06 ps. This suggests that there is a strong steric hindrance for the first process, since the geometric changes accompanying the trans-to-cis process require the displacement of membrane molecules (*vide infra*). The photoisomerization and the excited state decay of AZO-c are much faster than in AZO-t, both in vacuo and in several solvents,[24,27,38,44] because the slope along the torsional coordinate (CNNC dihedral) in the $S_1$ potential energy surface (PES) is much larger on the *cis* side than on the *trans* one. However, the difference in the lifetimes is here much larger than in other cases, thus indicating a more important role of the environment for AZO-t.

Analysing in more details the trans-to-cis process from Figure 4, we can observe a complete return of the population from $S_1$ to $S_0$ for both the active (the one finishing in the cis conformation) and inactive (the one returning to the trans conformer) trajectories, with a faster decay observed for the latter ones. On the other hand, no significant difference is observed for the cis-to-trans process, as both active (finishing in the trans isomer) and inactive (returning to the cis geometry) trajectories return to the ground state in the same time. Both these findings suggest that while for the trans-to-cis process the $S_1$ is populated for a fair amount of time, in the ps timescale, for the cis-to-trans the $S_1$ is only transiently populated, with an almost immediate decay to the ground state after photoexcitation.

The quantum yield for the trans-to-cis photoisomerization in DPPC of 0.24 is similar to the one obtained in similarly viscous solvents,[5] while there is an increase in quantum yield up to 0.65 for the cis-to-trans photoisomerization compared to gas phase (0.5) and similar to the one reported for ethylene glycol (0.67).[5] These results are consistent with literature on AZO embedded in different biological environments[36,37] and also when they are put in a self-assembly monolayer on gold.[38]

The reason why only the trans-to-cis process is affected by the presence of the environment is mainly due to the different interaction of the AZO-t and AZO-c with the DPPC membrane, as observed in the shift in the absorption spectra between the isomers. The observed differences in the electronic spectra and the decay parameters (Table 1) of AZO-t in vacuum compared to the ones in environment can in this respect be taken as a fingerprint of an interaction with the environment. Since however no shift is observed in the spectra for AZO-c and its decay constants are not different, an intrinsic proof is found that the membrane interacts differently with the two isomers. In fact, we observe a strong hydrophobic interaction between the phenyl rings of the azobenzene and the tails of the DPPC molecules only with the AZO-t isomer. On the contrary, the presence of a strong dipole moment for the AZO-c, its location close to the membrane surface (which is hydrophilic) and the distorted geometry of the isomer strongly decrease these interactions, nullifying the influence of the environment on the cis-to-trans process.

### 3.2.2 Photochemical pathway and isomerization mechanism

To rationalize our results, Figure 5 reports a schematic representation of the photodynamic pathways of the trans and cis azobenzene isomers in DPPC, as obtained from the ensemble of the QM/MM-SH trajectories. The reaction coordinate is an ensemble of torsional and bending modes, involving the CNNC and CNN internal coordinates.

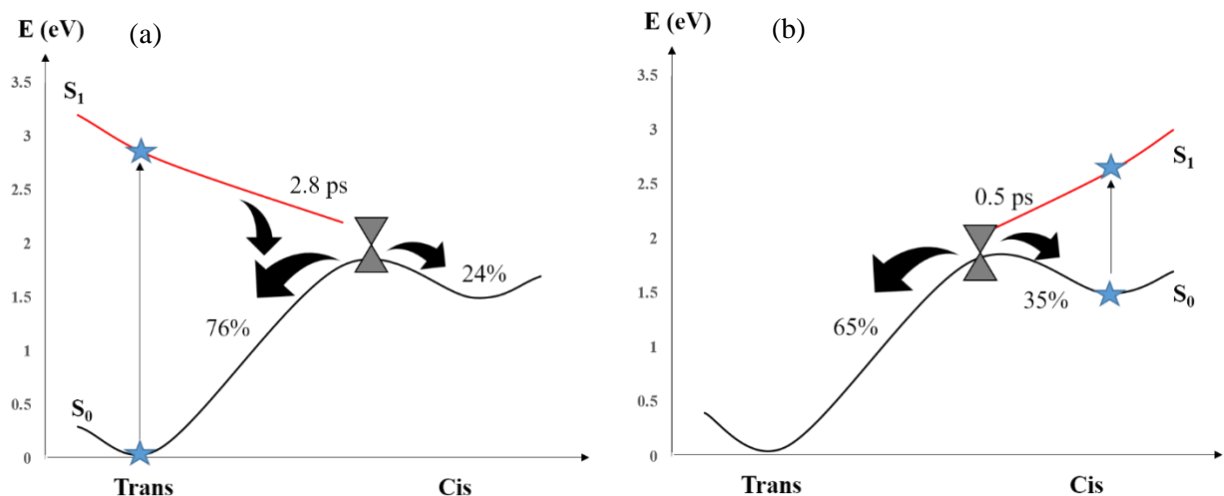

**Figure 5. Schematic representation of the photoisomerization of the azobenzene probe embedded in a DPPC lipid membrane at 323K as deduced from the present study, (a) for the trans-to-cis pathway and (b) for the cis-to-trans pathway. The $S_0$ ground state and S1(n–π*) state are shown explicitly (with calculated Franck–Condon energies indicated as a blue star). The effective reaction coordinate is a complex combination of internal modes, especially involving a concerted motion of the dihedrals. The conical intersection is also indicated schematically.**

An additional important factor to take into account (apart from the environment) is the difference in potential energy surface of the two isomers. In fact, if we analyze the $S_1$ potential energy as function of the CNNC torsion angle, we observe a much steeper curve on the cis side than on the trans side (Figure 5). This is also in agreement with different theoretical calculations.[25,39–41] As a result, the driving force for the isomerization is much stronger for the cis than the trans isomer, even in the presence of a hindered environment such a lipid membrane. Conversely, the shallow topology of the trans energy surface increases the exposure of the AZO-t to steric effects caused by the DPPC membrane.

If we have a closer look to what happens in the proximity of the conical intersection (CoIn) for the trans-to-cis process, we observe that the CNNC angle for the active trajectories has an average value of 107 degrees, with a very low energy difference ($S_1$-$S_0$) of 0.16 eV (Figure 6). All the active trajectories are found in a narrow energy gap and dihedral range of values. On the other hand, there is a much wider spread of the results considering the inactive trajectories, for which an average dihedral value of 119 degrees is observed together with a larger energy gap of 0.33 eV. This suggests that the inactive trajectories decay to the ground state before the intersection is reached. It tends to spend only a short time in the $S_1$ state and the jump to the ground state occurs at larger torsional values compared to the active ones (see discussion below). In fact, the presence of tightly packed lipid molecules in a DPPC membrane hinders the trans-to-cis photoisomerization, slowing down the process. Overall, the vast majority of the population (76%) is transferred back to the trans ground state geometry via unreactive trajectories at an early PES region, while of the ones reaching the conical intersection, only a minor part finishes in the cis ground state geometry, leading to an overall trans-to-cis photoisomerization yield of 24%.

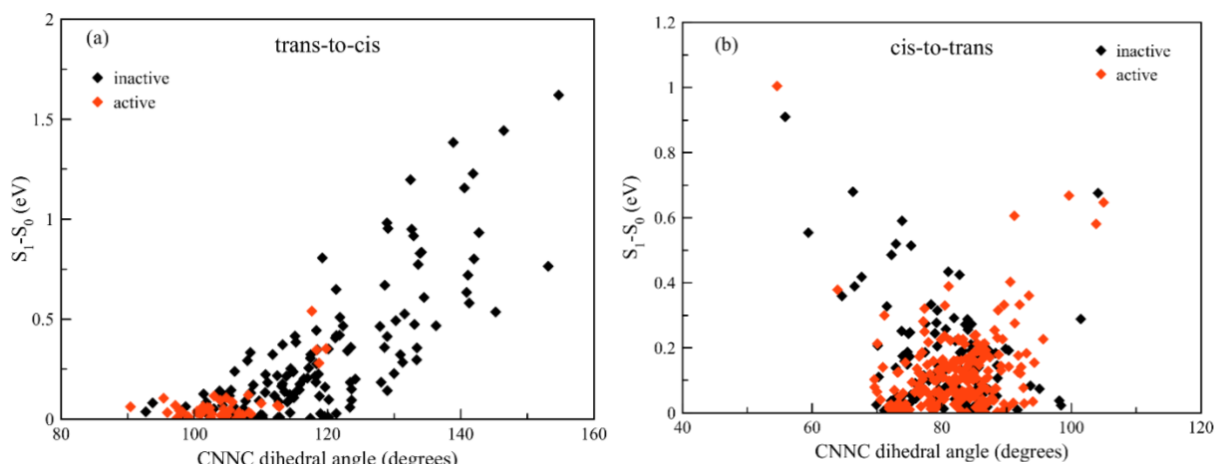

Figure 6. Energy distribution in function of the CNNC dihedral angle at the moment of the jump for the trans-to-cis (a) and cis-to-trans (b) photoisomerization.

An opposite situation is present for the cis-to-trans process. Now both active and inactive trajectories are localized in a small energy and dihedral ranges, with average values of 83 (81) degrees and 0.13 (0.16) eV for the active (inactive) ones. The simulations suggest thus a more effective inactive decay for AZO-t compared to AZO-c. All this is strong evidence of the presence of only one, fast reaction and/or decay mechanism, in which almost instantaneously after excitation of AZO-c, all trajectories reach the conical intersection, and from there decay either towards the trans or the cis isomer. Both populations of active and inactive trajectories present a $S_1$-$S_0$ energy difference at hops, and a similar CNNC dihedral distribution with a twisted geometry with average angle of 80 degrees. From this point, the population is transferred to the ground state with a prevalence of population ending in the trans ground state geometry (65%), leading to an overall cis-to-trans photoisomerization yield of 35%.

To gain more insight into the effect of the environment on the photoisomerization process, we consider the variation of the CNNC dihedral and CNN bond angles averaged over the active or the inactive trajectories running on the $S_1$ PES (Figure 7 and Figure S4 for the full time analysis).

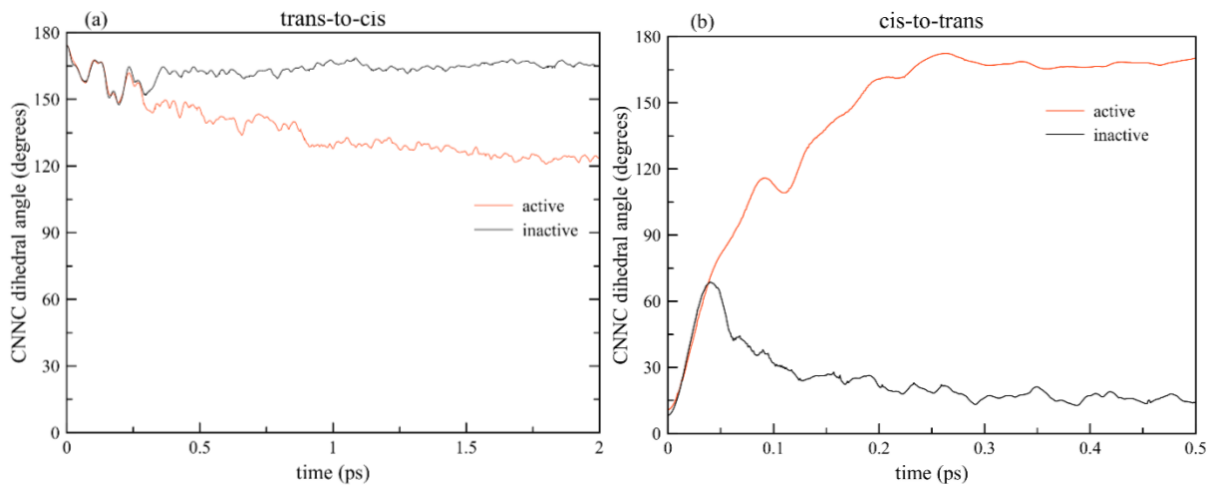
14

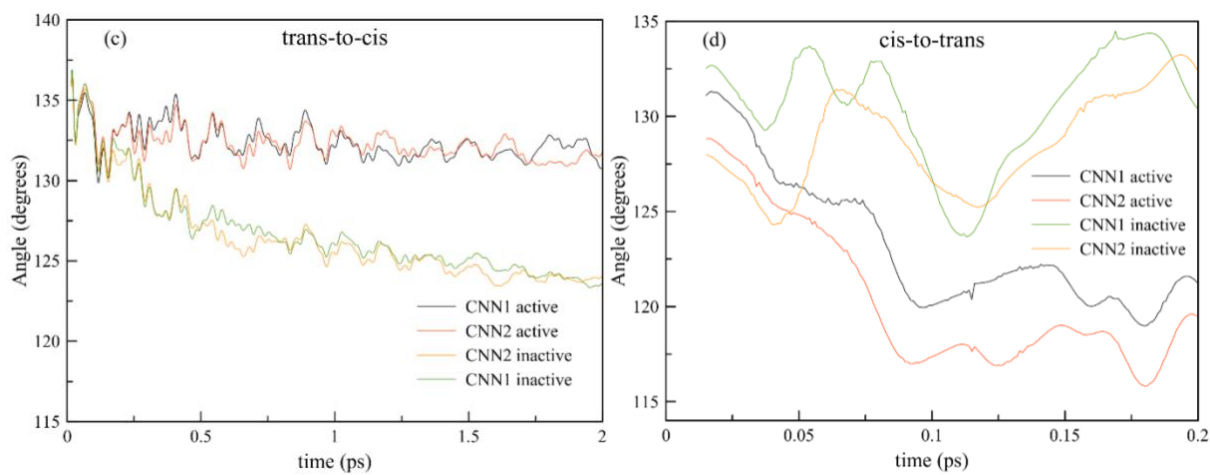

Figure 7. CNNC dihedral (a, b) and CNN angles (c, d) averaged over all trajectories running on the $S_1$ PES, for the trans-to-cis (a, c) and cis-to-trans (b, d) dynamics.

This helps to distinguish the dynamics in the excited state, which is crucial in determining decay rates and quantum yields, from the ground state one, where the focus is on the environmental effects on the relaxation towards the equilibrium geometries of the products. The plots clearly show that the dynamics following AZO-t excitation involves both the torsion of the N=N bond (CNNC coordinate) and the symmetric CNN bending vibration.[42] The latter is highly excited in the $S_1$ PES, because the equilibrium CNN in $S_1$ is much larger than in $S_0$ (132 vs. 117 degrees, in vacuo). Every time the CNN angles reach their upper turning point, the $S_0$ and $S_1$ PESs get closer and, with the help of a bit of torsion, decay to $S_0$ occurs, either for active or inactive processes.[43] This effect is more important in the trans-to-cis than in the cis-to-trans photoreaction, because the trans isomer shows a slower torsional motion compared to the cis isomer. Since it increases the decay rate without helping the isomerization, this effect decreases the quantum yield. While the CNNC dihedral angle is the most prominent parameter affecting the isomerization for both mechanisms, it can be remarked that a plateau is seen around 15 degrees (see Figure S4). Due to the presence of the DPPC membrane, the trans-to-cis photoisomerization it is not fully equilibrated as the lipid bilayer hinders the reach of the optimal dihedral angle for the cis isomer (indicating that a longer dynamic is needed to reach equilibrium). It can also be observed from the visual analysis of the different snapshots extracted from an active trajectory (Figure 8), in which the final orientation of AZO-c makes an angle of 65 degrees with the membrane normal (while it amounts to 80 degrees for an equilibrated system). This complex change of dihedral angle in different steps for the trans-to-cis isomerization is a strong evidence of a predominance of the torsional mechanism (scheme 1d in SI). This mechanism is also in agreement with the relative orientation of the AZO-t isomer in the membrane, with its long axis parallel to the lipid tails, favoring the torsion along the CNNC dihedral angle and maximizing the hydrophobic interactions with the membrane. On the other hand, the cis-to-trans isomerization exhibit an additional fast change of the CNN angles to 117 degrees for the active trajectories. This is a strong indication of a 'pedal like' mechanism (or Hula mechanism, scheme 1d in SI),[44] in which both dihedral and planar angles are involved in the isomerization process (Figure 8). We can exclude the presence of a pure inversion mechanism, since both the CNN angles change value, but at the same time the dihedral angle 'instantaneously' goes to 180 degree. This is also reported in literature[45] as 'volume conserving' mechanism, and the present case confirms the presence of a volume conservation due to the position of the AZO-c in the membrane, and the sub-ps decay observed. In fact, in the 'pedal like' mechanism, the movement is related to the displacement of the nitrogen atoms, which possess lone pairs which interact with the environment. Due to the position of the probe in the proximity to the lipid tails, this interaction is relatively weak. As a consequence, this cannot be the driving force for the isomerization. The movement is governed by entropy effects and the hydrophobicity of the phenyl ring of the AZO molecule. It should be noted that the final structure of the obtained AZO-t tends to be almost perpendicular to the membrane normal, while it was found to be parallel from MD equilibration. To reach the same equilibrium position as obtained from MD, longer time should thus be considered.

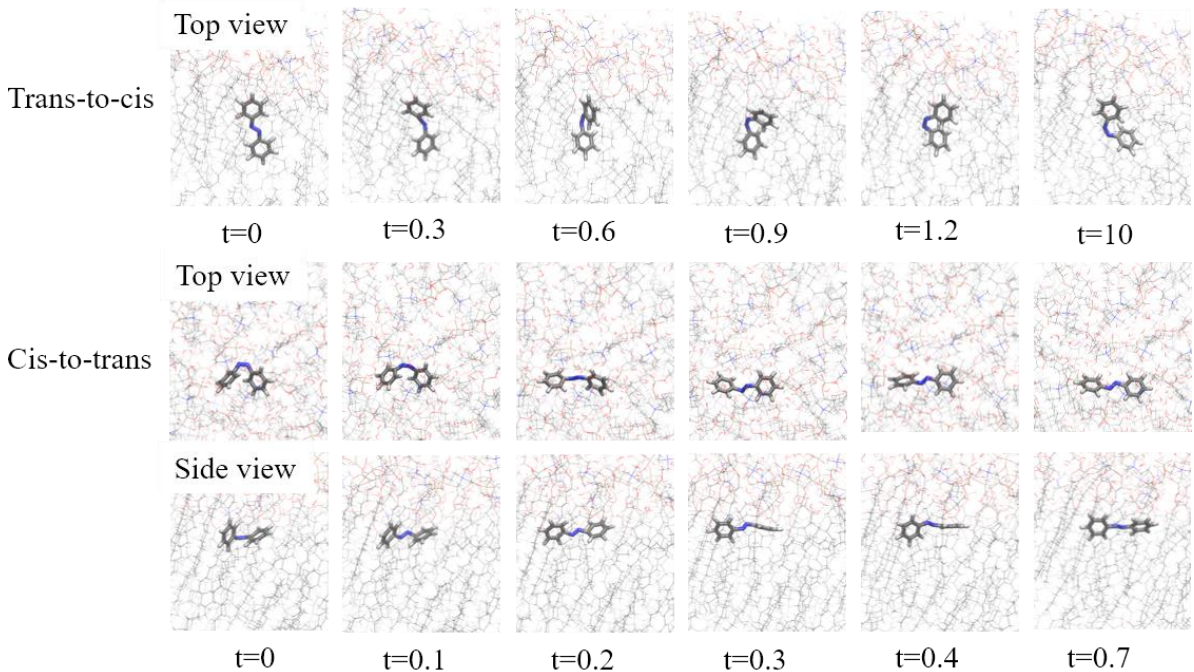

**Figure 8.** Selected snapshots along the reactive trajectory path for both trans-to-cis (upper panel) and cis-to-trans (bottom panel) processes. Time is reported in ps. We can distinguish the two suggested mechanisms: torsion for trans-to-cis and 'pedal-like' mechanism for cis-to-trans isomerization.

All the presented results largely depend on the fact that the PES from trans to CoIn and cis to CoIn are very different; for the trans isomer, there is almost no slope, while for the cis one the slope is very steep. These geometrical differences strongly affect the photoisomerization mechanism and the consequent decay times and quantum yields. In addition, the environment has a differential effect on the photoisomerization of AZO-c and AZO-t: almost negligible on the former, while for AZO-t it results in a slower isomerization due to stronger interaction of this isomer with the environment, through hydrophobic interactions with the tails of the DPPC membrane. Overall, AZO-t is more stable than AZO-c in the first excited state, and this hampers the initiation of the photoisomerization. This final aspect can be quantified by considering the difference in kinetic energy trends of the probe only (i.e. without the contribution from the environment) at the beginning of the process, just after absorption of one photon (Figure 9). From a starting energy of 0.9 eV, which is the equipartition value, we observe a strong, fast increase in energy for the AZO-c isomer up to 2.5 eV in the first 0.2 ps, which than slowly decays to a value of 1.5/1.7 eV depending on the active and inactive trajectories, respectively. On the other hand, a slow, steady and much less pronounced increase in energy is observed for AZO-t, reaching a maximum of 1.7 eV for the inactive trajectories and staying at a seemingly constant value of 1.3 eV for the active ones. If we compare this trend with the population analysis, we observe that the kinetic energy of AZO-c grows much faster than the decay to the ground state, confirming that the surplus of energy is here essential for the fast process observed, and is strongly related to the geometry of the PES for this isomer. The opposite holds true for AZO-t, for which the slow increase in energy directly relates to the flatter PES observed, together with the stronger interaction with the environment.

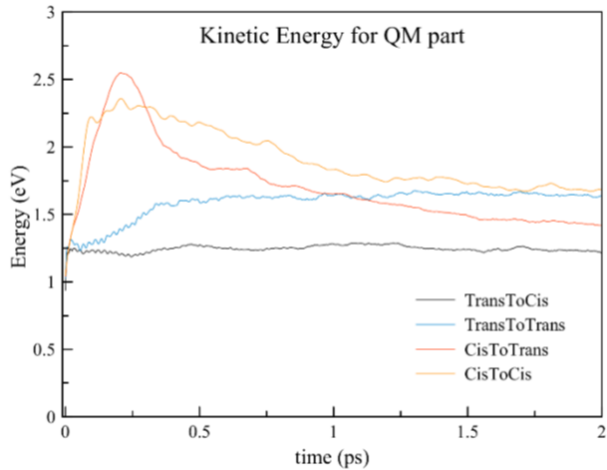

**Figure 9. Kinetic energy for the AZO probe for the first half of the process, in which the trajectories are on the $S_1$ excited state.**

### 3.2.3 Fluorescence properties

Time-resolved fluorescence experiments are very important to understand the emission properties of chromophores in different environments, giving at the same time information on the probe itself and the effect that the environment has on the emission properties. The emission rate averaged over all the trajectories is given by:

$$I_{tot}(t) = \frac{4}{3N_T \hbar^4 c^3} \sum_{k=1}^{N_T} \left[ \sum_{i=0}^{j-1} \omega_{ij}^3 \mu_{ij}^2 \right]^{(k)} \qquad (1)$$

Where $N_T$ is the total number of trajectories, $j$ is the current state of trajectory $k$ at time $t$, $\omega_{ij}$ the transition energy and $\mu_{ij}$ the associated transition dipole moment. Since fluorescence is occurring from $S_1$ to $S_0$, only these states are considered for the analysis. Histograms over $\lambda_{flu}$=hc/ω for given time intervals yield the transient fluorescent spectra $I(t,\lambda_{flu})$. The steady state fluorescence spectrum $I(\lambda_{flu})$ is obtained by time integration, and the fluorescence quantum yield is given by:

$$\Phi_F = \int_0^\infty I(\lambda_{flu})d\lambda_{flu} \qquad (2)$$

The computed fluorescence quantum yield is higher for the AZO-t than the AZO-c with values of $5 \cdot 10^{-6}$ and $2.11 \cdot 10^{-6}$ respectively, leading to a cis/trans ratio of 42%. The higher quantum yield for the AZO-t is related to its longer lifetime in the excited state, as well as to the effect of the environment, that enhances the forbidden $n \rightarrow \pi^*$ band. To gain a deeper insight into the dynamic effects, the transient fluorescence spectra should be analyzed. The fluorescence intensity as function of time is reported in Figure 10. The raw data are reported as a black curve for both isomers, and while we observe strong oscillations for the AZO-t before decaying after 2 ps, a much faster decay in less than 0.2 ps is observed for AZO-c. The raw data have been consecutively convoluted using a Gaussian function with fwhm=0.1 ps (which has been reported to be close to the experimental time resolution for the same probe in different solvents[24]) and fitted with a double exponential function $W_1 e^{-t/\tau_1} + W_2 e^{-t/\tau_2}$, which accurately describes the biexponential decay (see Table 2 for decay times and weight ratio). Interestingly, the first decay time is more than 16 times higher for AZO-t than AZO-c, although it is related to the intrinsic decay of the probe, while the second component is due to the presence of the environment, and is in line with what observed for the excited states lifetime. This discussion pinpoints an important difference between both isomers. Referring to its equilibrium position in the membrane, it can be said that the trans isomer is in an 'on' state when the polarization of light rather lies along the z-axis of the membrane. The AZO-c conformer can be seen as 'active' when the light is polarized along the membrane surface. The timescales and photophysics further on differ between AZO-t and AZO-c.

**Table 2. Exponential lifetimes and weights ratio obtained from the fitting procedure for both fluorescence intensities and anisotropies for the active trajectories. Times are reported in ps.**

|  | $\tau_1$ | $\tau_2$ | $W_2/W_1$ | $\tau_{an}$ | $r(\infty)$ |
|---|---|---|---|---|---|
| **AZO-t** | 1.13 | 3.75 | 0.83 | 0.18 | 0.36 |
| **AZO-c** | 0.03 | 0.03 | 1 | 0.15 | -0.06 |

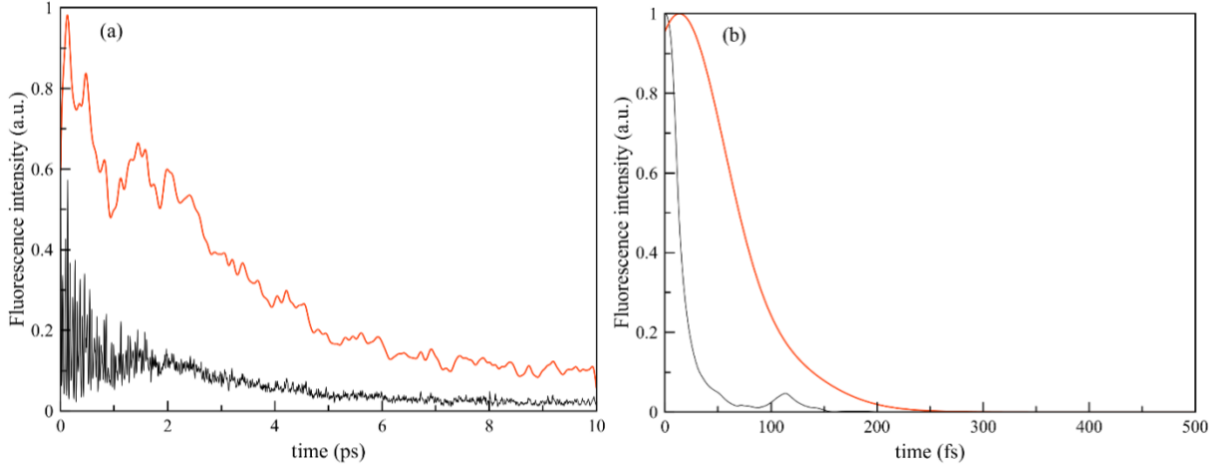

**Figure 10. Transient fluorescence spectra (black line) and its normalized convolution (red) for the trans-to-cis (a) and cis-to-trans (b) photoisomerization.**

Apart from the steady state and transient fluorescent spectra, the anisotropy and the decay time are often considered as markers to determine the effect of the environment. In particular, the time dependent fluorescence anisotropy ratio $r(t) = (I_\parallel - I_\perp)/(I_\parallel + 2I_\perp)$ is commonly used in experimental work and depends on the angle between the transition dipole moment of absorption at time 0 and of emission at time *t*. Immediately after excitation, with parallel transition dipole moment, a limiting value of *r*=2/5 is obtained, while with perpendicular transition dipole moments *r*= -1/5. When the same pair of excited states are involved in absorption and emission, only the geometrical relaxation and molecular rotation induced by the environment can change the direction of the transition dipole moment, and as a result the measurement of *r*(t) can reveal important information about the process.

It has been reported by both experiments and a computational studies[24] that the environment strongly affects both the fluorescence intensity and the anisotropy, especially when the viscosity of the environment is enhanced. To compute the parallel and perpendicular polarizations, equation (1) has been modified such as

$$I_\parallel(t) = \frac{4}{15 N_T \hbar^4 c^3} \sum_{k=1}^{N_T} \left[ \sum_{i=0}^{j-1} \omega_{ij}^3 \, \mu_{ij}^2 \left(1 + 2\cos^2 \beta_{ij}\right) \right]^{(k)} \quad (3)$$

$$I_\perp(t) = \frac{4}{15 N_T \hbar^4 c^3} \sum_{k=1}^{N_T} \left[ \sum_{i=0}^{j-1} \omega_{ij}^3 \, \mu_{ij}^2 \left(2 - \cos^2 \beta_{ij}\right) \right]^{(k)}, \quad (4)$$

where $\beta_{ij}$ denotes the angle between the transition dipole moment from state *j* to state *i* at times *0* and *t* for the *k*th trajectory. Moreover, as an anisotropic biological environment is present here, the depolarization is most likely not complete and an asymptotic *r(∞)* value has to be considered. To get a grip on it, the anisotropy data are fitted with the following function:

$$r(t) = r(\infty) + \left[\frac{2}{5} - r(\infty)\right]e^{-t/\tau_{an}} \quad (5)$$

The computed anisotropy decay is reported in Figure 11.

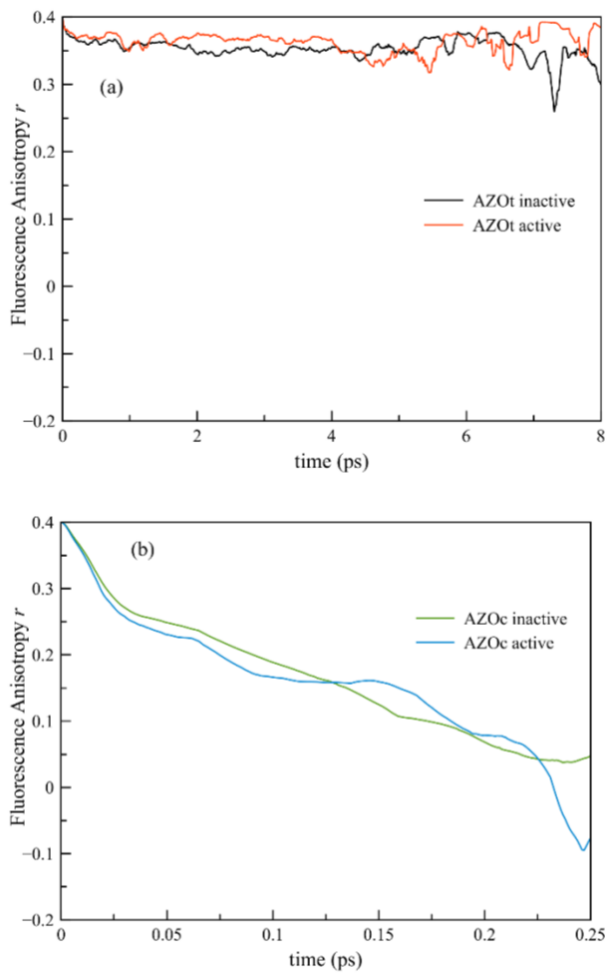

**Figure 11. Time evolution of the fluorescence anisotropy along the trajectories for trans-to-cis (a) and cis-to-trans (b) photoisomerization processes.**

Interestingly, we obtained a high r(∞) of 0.36 for AZO-t, which is close to the theoretical value of 0.4, while for AZO-c a negative value of -0.05 has been computed (data reported in Table 2).

From trans to cis, there is very limited intrinsic decay of the anisotropy; with a view on the time scales, any change of orientation has to come from a larger amplitude internal motion. From cis to trans, however, the opposite case is seen and a rather complete decay is observed along with a quite fast change of TDM orientation as AZO-c starts its pathway towards AZO-t. This translates into an effective difference between the two isomers when embedded in DPPC. In terms of anisotropy decay, the AZO-c isomer is the 'active' one. This result strongly suggests the value of AZO as fluorescent probe to determine membrane phase, since it will be possible to discriminate between the two isomers and having an 'on/off' probe. This effect was already reported in our previous study for similar probe, but now it is confirmed by more robust calculations.[15,16]

The initial depolarization rates, obtained as $(dr/dt)_{t=0} = [r(\infty) - 0.4]/\tau_{an}$ are very different between the two conformers, with a value of -0.2 ps$^{-1}$ for AZO-t and -3.1 ps$^{-1}$ for AZO-c. This means that the depolarization is more pronounced for the AZO-c, while for AZO-t its effect stops much earlier than AZO-c since $r(\infty)$ is closer to the theoretical value $r=2/5$. To understand these results, we must take into account the effect of the environment over the geometry of the conformers, and the orientation of the transition dipole moment (TDM). At equilibrium and in vacuum, TDM is negligible for AZO-t, but in the DPPC membrane the hydrophobic interactions between the environment and the probe distort the geometry of the conformer which is responsible for the oscillator strength for the $S_0$ to $S_1$ transition, which is weakly allowed. As reported in Section 1, the TDM direction is along the long molecular axis of the AZO-t, which keeps its alignment because of the environmental constraints. A completely opposite scenario is observed for the AZO-c conformer. In this case, the TDM direction makes an angle of 63 degrees with the N=N bond. The TDM of cis is known to rotate easily during the excited state dynamics and depends on the relatively easy rotation of the phenyl groups.[46] Moreover, once in the excited state, changes in internal coordinates take place very fast, thus enhancing the effect on anisotropy decay, which is much more pronounced than for the AZO-t, resulting in a much faster decay, and already after a few fs, the obtained intensities are minimal.

**Conclusions**

We presented here the computational study of the complete photoreaction mechanism of trans-to-cis and cis-to-trans photoisomerization of an azobenzene probe embedded in a biological environment, namely a DPPC membrane in its liquid disordered state. By means of classical molecular dynamics (MD) simulations coupled with non-adiabatic QM/MM surface hopping analyses (QM/MM-SH), we studied the effect of the environment on the photoisomerization mechanism, and found a dual mechanism which depends on the initial isomer of the azobenzene molecule. In fact, when considering the trans-to-cis isomerization, we observed a decay from the first excited state via a predominant torsional mechanism (with a minor contribution to the 'pedal-like' one), in which the molecule slowly converts into the cis isomer, leading a quantum yield (QY) of conversion of 24%. On the other hand, starting with the cis isomer we observed a sub-ps decay to the ground state, which occurs with a 'pedal-like' mechanism and is virtually not affected by the surrounding environment, leading to a total quantum yield of conversion of 65%.

To explain the difference in mechanism and QY we analyzed different key parameters, such as the geometry of the PES and the interaction with the environment. As a result, we found that AZO-c has a very steep PES in its first excited state, and coupled with negligible interactions with the environment, it explains the sub-ps decay and the high QY. On the other hand, AZO-t has strong hydrophobic interactions with the environment (i.e. the alkyl chains of the DPPC membrane molecules) which slow down the whole process. Together with the more flat PES in the first excited state, this explains the longer decay time observed, as well as the lower QY, since more trajectories can decay to the ground state before the conical intersection. These results are also supported by the difference in kinetic energy in the first half of the process.

The double photoisomerization mechanism was confirmed by both fluorescence spectra and anisotropy computations, the first with the longer decay time computed for the trans isomer compared to the cis, and the second by the much higher value of anisotropy for the trans compared to the cis, which is an indirect effect of the lack of interaction with the environment for the latter compared to the former isomer.

When the current results are combined with an experimentally oriented study on the photoisomerization of azobenzene in DPPC membranes, they can open potential applications in fluorescence and bioimaging sciences. As this study is to our knowledge one of the first which deals with the (simulation of) azobenzene probes in a lipid bilayer, we would like to encourage experimentally oriented groups to verify and eventually corroborate our findings.

**Acknowledgements**

S.O. thanks the Polish National Science Centre for funding (grant no. UMO-2018/31/D/ST4/01475 and UMO-2020-39-I-ST4-01446). This research was carried out with the support of the Interdisciplinary Center for Mathematical and Computational Modeling at the University of Warsaw (ICM UW) under grants no. G83-28 and GB80-24.

**Conflicts of interest**

There are no conflicts to declare.